\documentclass[11pt,conference,letterpaper]{IEEEtran}
\IEEEoverridecommandlockouts
\usepackage{cite}
\usepackage{amsmath,amssymb,amsfonts}
\usepackage{algorithmic}
\usepackage{graphicx}
\usepackage{textcomp}
\usepackage{xcolor}
\usepackage[hyphens]{url}
\usepackage{listings}
\usepackage{caption}
\graphicspath{{figures/}}

\def\BibTeX{{\rm B\kern-.05em{\sc i\kern-.025em b}\kern-.08em
		T\kern-.1667em\lower.7ex\hbox{E}\kern-.125emX}}

\makeatletter
\def\lst@makecaption{%
	\def\@captype{table}%
	\@makecaption
}
\makeatother

\begin{document}

\title{Please, do not decentralize the Internet with (permissionless) blockchains!
\thanks{
P.G. Lopez's work has been supported by the European Union’s Horizon 2020 research and innovation programme under the Grant Agreement No 825184 and by the Spanish government under project TIN2016-77836-C2-1-R. A. Datta's work has been supported by Singapore MoE Tier 1 Grant No 2018-T1-002-076.
}
}

\author{\IEEEauthorblockN{
Pedro Garcia Lopez\IEEEauthorrefmark{1},
Alberto Montresor\IEEEauthorrefmark{2},
Anwitaman Datta\IEEEauthorrefmark{3},
\IEEEauthorblockA{\IEEEauthorrefmark{1 }Universitat Rovira i Virgili, Tarragona, Spain \\
Email:  pedro.garcia@urv.cat}  
\IEEEauthorblockA{\IEEEauthorrefmark{2} University of Trento, Trento, Italy\\
Email: alberto.montresor@unitn.it}
\IEEEauthorblockA{\IEEEauthorrefmark{3} Nanyang Technological University, Singapore\\
Email: anwitaman@ntu.edu.sg}
}
}



\maketitle

\begin{abstract}

The old mantra of decentralizing the Internet is coming again with fanfare, this time around the blockchain technology hype. We have already seen a technology supposed to change the nature of the Internet: peer-to-peer. 
The reality is that peer-to-peer naming systems failed, peer-to-peer social networks failed, and yes, peer-to-peer  storage failed as well. 

In this paper, we will review the research on distributed systems in the last few years to identify the limits of \emph{open} peer-to-peer networks. We will address issues like system complexity, security and frailty, instability and performance. 

We will show how many of the aforementioned problems also apply to the recent breed of permissionless blockchain networks. The applicability of such systems to mature industrial applications is undermined by the same properties that make them so interesting for a libertarian audience: namely, their openness, their pseudo-anonymity and their unregulated cryptocurrencies. As such, we argue that permissionless blockchain networks are unsuitable to be the substrate for a decentralized Internet. 

Yet, there is still hope for more decentralization, albeit in a form somewhat limited with respect to the libertarian view of decentralized Internet: in cooperation rather than in competition with the superpowerful datacenters that dominate the world today. This is derived from the recent surge in interest in byzantine fault tolerance and permissioned blockchains, which opens the door to a world where use of trusted third parties is not the only way to arbitrate an ensemble of entities. The ability of establish trust through permissioned blockchains enables to move the control from the datacenters to the edge, truly realizing the promises of edge-centric computing.


\end{abstract}

\begin{IEEEkeywords}
decentralization, blockchains, peer-to-peer, edge-centric computing
\end{IEEEkeywords}

\section{Introduction}

The constant struggle between the forces of centralization and decentralization is inherent to the human nature and history. In the end, there is always a competition for power involving control and trust. In a centralized system, the participants must trust a central entity, which can effectively exercise control over them. In a decentralized system, both trust and control are spread between the participants.

In technology, we already witnessed the same competition.  In the origins of the Internet, cybernetics theorists foresaw a novel electronic frontier which would be decentralized, egalitarian, libertarian, and beyond the scope of traditional states. 
The birth of personal computing and the advent of the earliest virtual communities served to instantiate such a decentralized vision. However, even though the design of the Internet is inherently decentralized by nature, in practice, key functionalities rely on a very few service providers~\cite{giants}, who support and thus may effectively control the Internet.  

In this line, more than 75\% of the CDN market is controlled by three providers~\cite{cdns}, while five cloud service providers control around 60\% of the cloud market share~\cite{clouds}. Things are actually worse than what these numbers indicate individually, because a small set of entities in fact dominate several of these markets. For instance, Amazon alone controls almost 33\% of the cloud infrastructure market share, and through CloudFront, almost 40\% of the CDN market. Across different core Internet services there is a  similar concentration of power amongst a very few big players.   

While such a de facto centralization is likely a natural effect of market dynamics such as preferential attachment and a manifestation of power-law rather than a consequence of any technological bottlenecks,  it  is nevertheless a cause for concern due to a myriad of reasons, including the existence of a single point of failure, the structural risk of massively correlated failures, as well as privacy, censorship and anti-competition issues. 

In a previous wave of unsuccessful attempts, the peer-to-peer technology had seemed the key to providing a decentralizing Internet, because of an implicit belief that with a large user base, a proportionate amount of resources could be aggregated to scale  peer-to-peer services; and yet no single entity actually controls the service, which is inherently decentralized. 

With the advent of peer-to-peer cryptocurrencies, there has been a new wave of optimism, hoping that the incentive models of participation, resource contribution and consensus in crypto-currencies could provide a substrate for a decentralized Internet.

We argue that, unfortunately, despite addressing some of these issues in a self-contained manner, namely, making decentralized currencies feasible, 
decentralized permissionless blockchains remain ill-suited to be a substrate for a fully decentralized Internet.

The first goal of this article is to justify 
this claim
based on a historical review of existing literature in distributed systems. Particularly, we aim at showing the limits of open peer-to-peer systems and explain severe inherent problems (system complexity, security, fragility, performance) that preclude their indiscriminate adoption.

In particular, the incentive mechanisms that attract the resources for mining operations and thwart sybil attacks do not directly translate into a stable environment for creating other services such as storage or processing in a sustained manner, or in a stable priced manner which will be essential in the long term for both the resource providers and consumers. 

The second goal of this article is to present our vision for a decentralized Internet relying on a combination of permissioned blockchain networks, cloud computing, and edge-centric computing. As we can see in Figure 1, permissioned blockchain networks  could provide decentralized trust, edge-centric computing could ensure decentralized control, and cloud computing may become a true utility offering stable resources to users and organizations. 

\begin{figure}[t]
  \centering
  \includegraphics[width=0.45\textwidth]{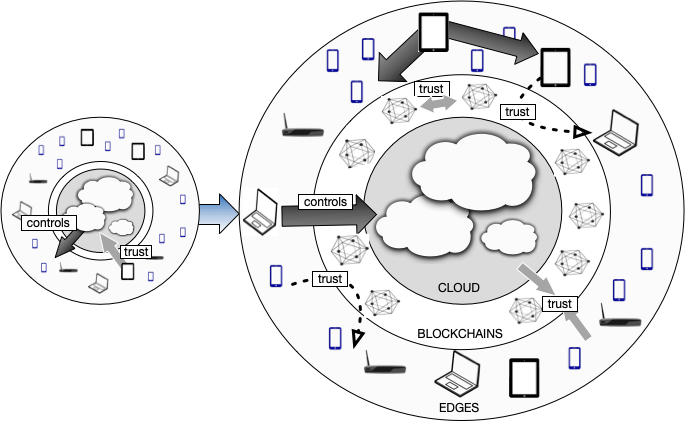}
  \caption{Centralized cloud model (left) versus Edge-centric Computing and Permissioned Blockchains (right).}
  \label{fig:architecture}
\end{figure}

While there is an important economic drive towards centralization, technologically speaking, we must not forget that ``everything is in the edge'': the humans that use the services, the fact that they are close to each other, the intelligent decisions and actuations that must be carried out. All these prompt an edge-centric architecture~\cite{edge}, consisting of a federation including not only big cloud datacenters, but also nano datacenters and personal devices. Control is moved to the edge, but leveraging the stable resources in the cloud. In this setup, while security and privacy are achieved through technologies such as end-to-end encryption, permissioned blockchains and smart contracts may enable trust between a medium-to-large collection of players, negotiations, validation of contractual obligations, (micro)payments among these entities.

We advocate for the ability to pool organizational and inter-organizational resources at the edge together, forming federations which possibly include public cloud services, and thus realizing decentralized applications or domain/industry specific islands of Internet, which do not suffer from the maladies of a de facto centralized Internet. In fact, we argue, it is feasible precisely with permissioned and private blockchains (something we have in fact started witnessing in their early avatars in several sectors, particularly including supply chains, banking sector, national blockchains), without the burdens of a permissionless blockchain. Furthermore, if the issue of interoperability of multiple blockchains is addressed properly, one can imagine multiple such decentralized groups which each rely on individual blockchains, forming amalgams (within as well as across domains/industries), to add to the degree of decentralization.

\section{History of P2P networks}

Peer-to-peer networks emerged in 1999 with the Napster service for sharing music files. Napster's architecture relied on a centralized index server for locating files that were later downloaded from peers. This network, as many other peer-to-peer networks afterwards, was completely open (permissionless in blockchain terminology). Anyone could enter the network with their PC and begin sharing files with others.

The system was finally closed due to copyright issues, but other alternatives rapidly emerged.
The first one was Gnutella, a completely decentralized overlay that relied on partial flooding for query messages. Gnutella is considered an unstructured overlay because nodes do not form any systematic topology and instead connect among themselves in a random manner. 

Gnutella, however, was slow and inefficient, and users abandoned it very quickly when better alternatives emerged. Superpeer overlays solved the problem including a layer with more stable peers that boosted the overall performance. Many systems like Kazaa, eMule, eDonkey or even Skype relied on such superpeer architecture.

At the same time, peer-to-peer research sprouted with very interesting contributions, e.g., gossip based protocols for scalable group communication and DHTs that have inspired the design of key-value stores and the data model for many NoSQL systems.

\subsection{Decentralizing Internet with DHTs}

The first decade of this century witnessed the emergence of the concept of structured overlays, or Distributed Hash Tables (DHTs), with numerous proposals such as CAN~\cite{ratnasamy2001scalable}, Chord~\cite{stoica2001chord}, Pastry~\cite{rowstron2001pastry} and Kademlia~\cite{maymounkov2002kademlia}. 

In the academic world, everything could be built on top of a DHT. Researchers proposed decentralized naming systems~\cite{pappas2006comparative}, decentralized search mechanism~\cite{harren2002complex},  decentralized notification systems~\cite{rowstron2001scribe}, decentralized storage~\cite{rowstron2001storage}, decentralized file systems~\cite{dabek2001wide}, or decentralized mail notification~\cite{mislove2003post} among many others.

There were so many DHTs that some researchers presented a common API for structured overlays to promote interoperability~\cite{dabek2003towards}.  But they also noticed scarce adoption of DHTs and proposed a public DHT as an universal substrate for applications~\cite{karp2004spurring}.

In many cases, DHT academic designs were fragile and complex to implement in the real world. But finally, DHTs were adopted in systems like eDonkey KAD and BitTorrent DHTs, with millions of users.

Some studies analyzed those massive DHTs~\cite{steiner2009long}~\cite{wang2008attacking} and outlined their major challenges. In particular, open networks where peers can assign their identities are prone to Sybil attacks~\cite{douceur2002sybil}. In a Sybil attack, the idea is to impersonate thousands of identifiers with a few powerful nodes in those networks.

In general, these open overlays were scalable and provided a reliable service, but they were unstable due to churn and connectivity problems, and suffered some performance problems. For example, Jim{\'e}nez et al. reported that  lookups were performed within 5 seconds 90\% of the time in Emule's Kad, but the median lookup time was around a minute in both BitTorrent DHTs\cite{jimenez2009connectivity}.

The structural fragility of those open networks and their performance issues precluded their usage in practice as global substrate for different kinds of applications. DHT research, however, turned out to be highly beneficial in the design of reliable cloud infrastructure in the following years.

In any case, the idea of decentralizing the Internet also inspired several projects that finally went to the real world. To name a few, we mention Diaspora, as a decentralized social network (2010), Freedombox, a personal safe appliance (2010), arkOS (2014), Yacy, a decentralized search engine (2003), and Wuala, a decentralized storage system (2008).

The major motivation for all those peer-to-peer alternatives was privacy and the idea of retaking control of user’s data. In the era of personal computers, users were enabled to decide where their data was stored and processed.  But with the advent of cloud (based) services like Google, Amazon, Dropbox or Facebook, a selected few of these service providers have  access to a disproportionately vast portion of the data. As a direct consequence, most online services rely on a very small set of entities, which is undesirable both in terms of user privacy and censorship resilience, but also in terms of being single points of failures. 

\subsection{Limitations and problems of open P2P networks}

In the end, peer-to-peer technologies faced strong challenges and problems that finally precluded their massive adoption and survival beyond in best-effort, file-sharing applications:

\textbf{Problem 1: Free Riding and incentives}. 
In general, users do not donate their computing, storage and bandwidth resources for altruist reasons (exceptions like SETI@ home exist). Normally, users minimize their time connected until obtaining what they want (video or music files). This is called \emph{free riding}, an issue that was extensively reported in the Gnutella overlay~\cite{adar2000free}.

BitTorrent mitigated the free riding problem by  designing the protocol including incentives (tit-for-tat). If peers do not contribute, others would not reciprocate. But again, collaboration is only enforced during the download process.
Unfortunately, users have little incentive to maintain an open infrastructure like a DHT. In eDonkey and BitTorrent DHTs, there is high churn (nodes entering and leaving the network) because of this lack of incentives.

Incentives can also provoke the opposite effect. If there are many incentives (economic ones), powerful economic players could then take over the network to grasp those incentives. We discuss this point further in section 3.

\textbf{Problem 2: Performance problems due to instability, heterogeneity and churn}. 
Because of Problem 1, P2P networks show high heterogeneity and high degrees of churn. To maintain the service these protocols must be fault-tolerant and self-adjusting, but this can cause performance problems and latency. When one needs any kind of guaranteed quality of service with stringent constraints such as millisecond response time, or to process a large volume of transactions, stable cloud servers have no rival in P2P networks.

\textbf{Problem 3: Security and fragility of open networks and peers}. 
In the absence of a centralized bootstrap process, any node can enter the network without control. In structured overlays, this also means that they can assign themselves identifiers, and that the overall network is vulnerable to sybil attacks. Massive identity problems were reported in eMule KAD and in Bittorrent DHTs~\cite{steiner2009long}.
Furthermore, peer nodes are more vulnerable to attacks by third parties. Whereas data centers have security measures, uncontrolled PCs and user devices may be exposed by software or hardware vulnerabilities. Even corporate user devices like WD MyCloud~\cite{wdmycloud} reported massive vulnerabilities that enabled third-parties to access the data and open backdoors. Furthermore, the P2P network itself can act as a potent channel for spreading malwares.

\textbf{Problem 4: System complexity and maintenance}. 
Last but not least, programming and maintaining peer-to-peer software system is a complex task.  Programming distributed systems that must be fault-tolerant, self-adjusting, and scalable is a challenging task. Furthermore, open protocols may have a myriad of faulty implementations provoking continuous problems. BitTorrent's creator Bram Cohen complained about the low quality of many BT clients and considered releasing a closed protocol and client for its novel streaming service.  Wuala discontinued the P2P storage and moved completely to cloud storage in part motivated by software complexity and instability. 

Apart from these problems, some researchers discovered that for many networks, $O(1)$ networks  or one-hop overlays~\cite{ramasubramanian2004beehive}  were feasible. In particular,~\cite{gupta2003one} demonstrated that for networks between 10K and 100K it is possible to have full membership routing information and provide one-hop routing. If the overlay is relatively stable like a corporate network, then $O(1)$ routing  and  full membership is the right decision instead of maintaining routing tables and suffering multi-hop lookups.

The research on DHTs, gossip-based protocols, and O(1) overlays clearly paved the way to distributed cloud-based key-value stores and NoSQL systems. For example, Amazon Dynamo ~\cite{decandia2007dynamo} leverages structured overlays (Chord, Pastry), gossip-based communication, and one-hop overlays. Dynamo, like other scalable NoSQL systems, is built on top of stable and trusted cloud servers which avoids problems 1-4 but also simplifies considerably the software architecture and boosts the overall performance.

Even if content-addressable storage and object storage devices already existed, DHTs also influenced the creation of cloud object storage systems following the same principles of key-value stores but for big storage blobs. In this line, Azure Storage~\cite{calder2011windows} is inspired by Amazon Dynamo and Cassandra, among others. 
Cloud Object Stores like Amazon S3 or Azure Storage were massively adopted by large applications like Dropbox, Netflix and many others. Amazon S3 claimed 2 trillions objects already in 2013. 

\subsection{Decline of P2P}

Even if many of the innovations from P2P research were incorporated in and inspired the design of humongous datacenters, the advent of cloud computing also killed the P2P star. Research on peer-to-peer dried up and the major conferences dedicated to the theme ceased, IPTPS in 2010 and IEEE P2P in 2015. 

BitTorrent accounted for 35\% of Internet traffic in 2004, but only 3\% in 2018. Diaspora only reached one million users, and many projects were abandoned: arkOS, OpenDHT, or Yacy among many others. Wuala abandoned their P2P storage mechanisms and focused entirely in cloud storage. Later on it was acquired by Seagate. 
Skype changed its superpeer-based approach with data center based stable servers.

And traffic in the major P2P file-sharing sites like eDonkey or BitTorrent notoriously declined, in part because of cloud-based legal sites such as Netflix, Amazon Prime Video and so on. 

Today, the world is clearly dominated by GAFAs (Google, Amazon, Facebook, Apple). All major global services are cloud-based. It seemed that P2P had died or become marginal, but then cryptocurrencies got mainstream. We begin to hear again the old drums about decentralizing the Internet and fight the Big Brother.

\section{Bitcoin and the blockchain hype}

The original Bitcoin paper written by the obscure Satoshi Nakamoto appeared
in 2008~\cite{nakamoto2008Bitcoin}. Nakamoto proved to the world that an
ingenious combination of existing technologies and economic incentives could
achieve a long-sought goal: a completely decentralized virtual currency,
without intermediaries, open to everybody but at the same time capable of
solving problems such as double spending and sybil attacks.

\subsection{The basics}

Bitcoin and other open blockchains provide the abstraction of a single
distributed \emph{ledger} of transactions. In their most basic forms,
transactions look like `payer $a$ sends $x$ bitcoins to payee $b$' requests,
although more complex types of transactions exist. Transactions are collected
into blocks, each block containing a hash of the previous one up to the first
(\emph{genesis}) block, effectively creating a \emph{blockchain} (hence the
name).

In order to achieve the decentralized verification of every transaction, the
entire blockchain is replicated among all nodes (peers) forming the network.
Participation is open to everybody; peers may join freely just by contacting
some open bootstraps. Nodes are organized in an unstructured topology where
each peer is connected to a large number of other peers, randomly. Transactions
generated by nodes are disseminated through the entire network and verified
independently by each node, to intercept and avoid double spending.

The chain is extended through a process called \emph{mining}. Nodes
collect transactions into blocks, and then try to make a new block
accepted by everybody else as the next block in the chain by providing a
\emph{proof-of-work} through the solution of a cryptographic puzzle: the miner
looks for a random number called \emph{nonce}, such that when the block
content and the hash of the previous one is hashed along with the nonce, the
result is numerically smaller than a given \emph{difficulty target}. The
difficulty target is periodically adjusted in such a way that a new block is
generated every 10 minutes.

As mining is computationally (and economically) expensive, every time a miner creates a new block, the miner is rewarded in two ways: through the creation of new bitcoins and through transaction fees that may be offered by who requests a transaction to be executed. This incentive scheme is supposed to keep the network alive and is a major departure from existing peer-to-peer networks where free-riding is predominant.

The proof-of-work mechanism is paramount to the success of the protocol: it
enables leader election, provides a (non-final) byzantine fault-tolerant
consensus service, guarantees the immutability of the chain and offers a
defense against sybil attacks. The value assigned to a bitcoin in an open market provides the ultimate incentive for miners (to spend on the capital investment and operational costs). The creator of the next block is elected
randomly by the process of mining the nonce. Given the probabilistic nature of
the process, the blockchain may occasionally \emph{fork}: the chain may be
extended by distinct blocks. As nodes are incentivized to extend the longest
fork, such ephemeral forks quickly disappear, reaching a (delayed) consensus.
The choice of the longest chain is also useful to guarantee immutability: since
blocks are chained through their hashes, modifying the content of a block
requires to re-compute the proof-of-work for that block and for any block that
follows, obtaining a chain longer than the official one; a feat possible only
if the attacker possesses more than half of the computing power. Having multiple (anonymous) identities, as in sybil attacks, is thus useless.

\subsection{The hype}

The growth of interest in Bitcoin had a slow start. Initially discussed among security researchers and practitioners, the first major users of Bitcoin were black markets, such as Silk Road. The first official Bitcoin exchanges were created in 2011. The exchange rate grew slowly at the beginning (\$0.30 in 2011, \$5.27 in 2012, \$13.30 in 2013), but things started to definitely change in 2014, when the rate reached \$770, to culminate in December 2017 when the Bitcoin price peaked (till the time of finalizing this paper in April 2019) at $\sim$\$19,783. 

The increased interest has provoked a gold rush around Bitcoin mining, with specialized BitFarms (many of them Chinese) that use specialized hardware and consume tons of electricity. According to the Economist~\cite{economist}, the Bitcoin energy consumption peaked at 70TWh in 2018, which is roughly what a country like Austria consumes.

Cryptocurrencies and blockchain technologies have also attracted a lot of investments. In particular, \emph{initial coin offerings} (ICOs) are popular. According to CoinSchedule, more than one thousand ICOs have been launched in 2018---almost a three-fold increase with respect to 2017~\cite{economist}--raising more than \$21 billions.

To give some examples, Ethereum raised \$18m in 2014, Block.one raised \$4bn for the EOS blockchain, and Telegram raised \$1.7bn for their cryptocurrency. Because a rising tide lifts all boats, many “old” peer-to-peer companies are also reviving. In this line, the blockchain startup Tron bought the BitTorrent company for \$140m, and veteran IPFS has revived with FileCoin. FileCoin leverages IPFS to propose again a decentralized storage network combining blockchain with proof-of-space and user’s resources. FileCoin's ICO recently received \$200 million from accredited investors.

Blockchains are now reviving the idea of decentralizing the Internet, but now with a lot of investors' money. Wired recently reported in a number of decentralized applications that try to replace cloud ones like DTube (decentralized YouTube), Graphite Docs (decentralized Google Docs), Hearth (decentralized Dropbox and Web hosting service), Props (decentralized video streaming), FileCoin (decentralized storage), and  OpenBazaar (decentralized eBay) among others~\cite{wired}.

\subsection{Problems of permissionless blockchain networks}

The current investor euphoria around blockchains is inherently tied to lack of knowledge in complex technologies. We aim to shed some light on this issue and to clarify when it is appropriate to use decentralized technologies like blockchains, and when it is unnecessary or even completely absurd.

First of all, any decentralized peer-to-peer overlay with open membership still suffers the four problems stated before. This also applies to open blockchain overlays like Bitcoin and Ethereum.

\textbf{Problem 1: Free riding and incentives}. 
Open P2P networks must offer incentives so that users stay connected and contribute with their resources (computing, storage, bandwidth). As stated in the previous section, if the overlay does not provide enough incentives, the network can suffer free riding, high instability and churn.

On the contrary, when strong incentives are at play, external economic actors may appear and even disrupt the network. In the original Bitcoin paper \cite{nakamoto2008Bitcoin}, incentives are economic and clear: \textit{``By convention, the first transaction in a block is a special transaction that starts a new coin owned by the creator of the block. This adds an incentive for nodes to support the network, and provides a way to initially distribute coins into circulation, since there is no central authority to issue them."} The author also states that peers vote with their CPU power, and then \textit{``it quickly becomes computationally impractical for an attacker if honest nodes control a majority of CPU power"}. 

As the economic value of bitcoins raised in the last years, these incentives
were really attractive for many companies. Huge commercial BitFarms with
specialized hardware emerged to mine bitcoins. In 2013 six mining pools
controlled 75\% of overall Bitcoin hashing power. Nowadays it is almost
impossible for a normal user to mine bitcoins with a normal desktop computer.
Some recent research work~\cite{eyal2018majority} indicates that the
incentive mechanism of Bitcoin is furthermore flawed. They present an attack where a
minority colluding pool can obtain more revenue than the pool’s fair share.

These economic factors are compelling  normal users to quit/stay out from participating in the mining process. In a broadcast
network where all nodes validate transactions, and where the history of
transactions grows, each node requires more bandwidth, more storage, and more
computing power to cope with the flow.

To avoid network shrinkage due to nodes being expelled by resource limits, some
networks are retagging nodes as “light nodes” but they still count them in the
global network size metrics. For example, in Ethereum, a sharding architecture
has been introduced to improve performance which distinguishes between full and
light clients. Full clients validate transactions whereas light clients do not
participate in transactions. 

Furthermore, given the volatility of cryptocurrency valuations, this leads to a situation significantly worse than usual commercial cloud based services, by causing great pricing instability and uncertainty both for the service consumers, and also the resource contributors.

\textbf{Problem 2: Performance problems}. One of the most severe problems of
blockchains is performance. While VISA is processing 24,000 transactions per
second, Bitcoin can process between 3.3 and 7 transactions per second, and
Ethereum around 15 per second. This is the consequence of a large unstructured
broadcast network where all nodes validate transactions. VISA can rely on a
smaller pool of cloud servers that partition traffic and handle tons of
transactions per second.

In this line, Ethereum's creator Vitalik Buterin proposed the \emph{scalability
trilemma} that states that a blockchain technology can only address two of the
three challenges: scalability, decentralization, and security~\cite{trilemma}.
For Buterin, scalability is defined as being able to process $O(n) > O(c)$
transactions, where $c$ refers to computational resources (including
computation, bandwidth and storage) available at each node, and $n$ refers to
the total size of the ecosystem.

In fact, although most cryptocurrencies and blockchains claim
decentralization, many of the new and existing networks are proposing more
centralized designs to increase the overall performance. The so-called layer 2
or off-chain solutions like Lightning network (Bitcoin), Plasma (Ethereum) or
EOS follow this trend. In these cases, transactions are processed by a much
smaller set of peers (outside the core network) to increase performance.

An important strength of permissionless blockchains is that everybody can join, irrespective of who they are. This is a big departure from traditional byzantine fault tolerance, where the number of entities participating in the protocol is limited. 
Openness come with a price, though. To avoid sybil attacks, Bitcoin adopted proof-of-work, resulting in a huge waste of energy resources. Alternative approaches based on proof-of-X, where X could be stake, space, activity, etc. seem not be able to fully address this problem~\cite{proof-of-stake} so far. 

Our criticism on performance is clearly associated to the current breed of permissionless blockchains;  it is  possible that future  blockchains improve their overall performance.  But they are not designed as a general purpose computing and storage platform for third-party applications. So the basic question will be: will the additional price tag to be payed to be open and decentralized be ever small enough to make permissionless blockchains competitive against cloud implementations?



A completely open network of thousands of heterogeneous nodes is a serious burden for performance, and in direct conflict with commercial uses of the technology requiring efficient transactions and transfers. 


\textbf{Problem 3: Security and fragility of open networks and peers} Security
is another very relevant issue in a permissionless or open blockchain networks
like Bitcoin or Ethereum. Attackers in this case have economic incentives to
break the system. In this line, Bitcoin has been the target of many attacks and
hacks as reported in~\cite{hacks}. Not only users lost their bitcoins in those
attacks, but even big Bitcoin exchange sites like Mt. Gox, Bitfinex or BitStamp
reported major intrusions stealing them of millions worth in bitcoins.

Ethereum has also its own history of attacks and hacks. In 2016, the DAO
(Decentralized Autonomous Organization) raised US\$150 millions to fund the
project, but a hacker found a bug in the code that enabled him to make off with
US\$50 millions in ether currency. This even provoked a fork creating two
networks: Ethereum Classic and Ethereum. From 2016, Ethereum suffered a second
fork and they are still increasing their DDoS protection and security
countermeasures to protect the network and code.

The decentralized apps (DApps) built on top of smart contracts are another
important source of vulnerabilities (even inside sandboxed VMs). For example, in 2017, a game called CryptoKitties (built using smart
contracts) went viral and traffic on Ethereum’s network rose sixfold provoking
the failure of many transactions.


In any case, while smart contracts enable all sorts of experiments and even the
creation of new cryptocurrencies that leverage the Ethereum network,
Blockchains are still an open experiment with many challenges to solve.

\textbf{Problem 4: System complexity and maintenance}.
The last important problem is system complexity and software maintenance/evolution. The code of permissionless blockchains like Bitcoin or Ethereum is
huge and complex, and is constantly evolving to cope with software
updates, security issues and new features.

It is clear that decentralized systems are clearly more complex to develop and
maintain than their centralized counterparts. As we already explained, this was
in part a strong reason to abandon peer-to-peer storage and focus on cloud
storage by companies like Wuala.

In Ethereum, the software is even more complex because of the EVM and Smart
Contracts. Using the Solidity language it is possible to develop code and
create DApps that run on top of the Ethereum network. DApps are small scripts
that can transfer currency and connect users. They are good at coordinating
lots of computers to perform tasks in exchange for currency without any central
control. For example, with Golem Dapp, users can sell their machine’s unused
computing power or buy it from others.

DApps and smart contracts, however, are complex to code and fragile. DApps can
call into the public methods of other programs and it is very difficult to
write flawless code. For example, a recent vulnerability in the Parity Wallet
Dapp allowed hackers to steal \$30 million in ether. And storing state in a
smart contract may be extremely expensive due to the inherent costs of the
Ethereum network. Ultimately, the incentive mechanisms of cryptocurrencies work well in a self-contained manner, yet the incentives do not translate to, and even hinder, other applications.

\subsection{Permissionless blockchains are not the right way}

\emph{Permissionless blockchain networks are not going to be the substrate of a decentralized Internet}. They are slow and expensive beasts that are unable to sustain massive global Internet services. Scalable distributed services require a combination of stable and fault-tolerant servers, content-distribution networks, and shared-nothing partitioned architectures (like cloud object stores) that do not fit with open peer-to-peer networks (less with open broadcast peer-to-peer blockchains).

The consolidation of a core backbone of commercial stable servers to sustain the cryptocurrency, along with the expensive and volatile cost of transactions, will drive out from the network any efforts to use it as a cheap repository for third-party applications. It is cheaper (and safer) to use cloud standard services with stable pricing models. Furthermore, the unnecessary entanglement of a volatile currency to the services just increase the uncertainties and risks associated to the platform.

\section{Permissioned blockchains to the rescue}

There is a completely different audience that demonstrated an extensive
interest for blockchains: surprisingly, the same groups that were supposed to be made obsolete by (decentralized) blockchains, namely companies, banks, mortgage registries, governments and so on.

Their attention is not focused on permissionless blockchains, though. Instead,
they promote the development of \emph{permissioned} blockchains, operated by an
ensemble of known entities such as members of a consortium or stakeholders in a
given business context. 

Unlike permissionless ones, permissioned blockchains have means to authenticate the nodes that control and update the shared state and to authorize who can issue transactions. Yet, the members of such blockchains do not necessarily trust each other. They use distributed mechanisms that act in lieu of a trusted and dependable third party, maintaining the shared state and mediating the exchanges. Very often, permissioned blockchains offer a secure computing engine as well, capable to execute arbitrary tasks in the form of smart contracts.

Once blockchains are disentangled from cryptocurrencies (and the associated incentive mechanisms), an old problem resurfaces, which has kept busy ranks of researchers for over two decades: byzantine fault tolerance.
In fact, the advent of permissioned blockchains has given new life to research on practical solutions to problems like consensus, state replication and epidemic broadcast, in environments where network connectivity is uncertain, nodes may crash or become subverted by an adversary, and interactions among nodes are inherently asynchronous.

Many groups of researchers are competing to develop more and more efficient and secure permissioned blockchains, such as Hyperledger Fabric~\cite{hyperledger}, Tendermint, Symbiont, R3 Corda, just to mention a few. 

It is outside the scope of this paper to describe all of them or to identify a clear winner, but we use Hyperledger Fabric as a reference architecture for permissioned blockchains, given the extensive support its umbrella project (Hyperledger) receives from the Linux Foundation and from companies like IBM and Intel.

The software architecture of Hyperledger Fabric is modular and enables different pluggable components. It avoids costly proof-of-work by using different consensus algorithms such as crash fault-tolerant (CFT) or byzantine fault tolerant (BFT) protocols, the latter based on BFT-SMaRt~\cite{bftsmart}.
Additional needed services can be exploited and configured as well, such as distributed membership (associating entities in the network with cryptographic identities), gossip-based dissemination services, and smart contracts engines that enable the execution---in sandboxed environments --- of \emph{chaincode} software written in standard programming languages.

One distinguishing aspect of Hyperledger Fabric is that consensus or replication can be configured between a subset of the nodes of the network, unlike traditional broadcast networks (like Bitcoin or Ethereum) where all nodes must participate in all transactions.

 In fact, permissioned blockchains are now thriving as commercial cloud services.  \emph{Blockchain-as-a-service} (BaaS) offerings have rapidly grown to include tech industry's leaders such as Oracle, Amazon, IBM, or Microsoft among others.  For example, Oracle's BlockChain Cloud, built on Hyperledger Fabric, can provide blockchain services over reliable cloud resources, and thus guaranteeing performance and security. Amazon is also offering scalable cloud-based blockchain services like Amazon Quantum Ledger Database (QLDB) and Amazon Managed Blockchain. 

In the next section we will explain why we believe that permissioned blockchain stacks may become essential tools for a decentralized Internet.

\section{Edge-centric Computing and Permissioned Blockchains: A proposal for decentralizing the Internet}

In a previous paper, we proposed \emph{edge-centric computing}
as an evolutionary paradigm that will push the frontier of computing
applications, data and services away from centralized nodes to the periphery
of the network~\cite{edge}. 
In the edge-centric vision, ``everything is in the edge'': not only the humans that use the services, but also the activities that put them in contact in the first place. \textbf{Control must be at the edge}, because both the information needed to take decisions and the effects that the decisions may have are located there. 
Other edge-computing and fog-computing approaches just focus on computation close to the data, where a large amount of pre-processing is done before dumping the data into the cloud. In that scenario, edges are mere appendices of the Cloud, which effectively controls them completely.

But our edge-centric computing proposal missed an important point: \textbf{decentralized trust}. Our thesis is that permissioned blokchains combined with decentralized identity mechanisms are the required building blocks for the decentralized Internet of the Future.

When interacting devices belong to different administrative domains, each of them could be an independent and potentially opportunistic player; so, in order to cooperate effectively, some level of trust must be gained. 

In the real world, trust is normally established by central authorities, from environmental agencies to food quality consortiums to government-based services. Such approach can and is replicated in the online world, where centralized services put their weight to guarantee an adequate level of trust.

Yet, with the increase in the number of operations that are moved from the physical world to the virtual one, such centralized approach poses both technological and political concerns. First of all, modern services are data-intensive and latency-sensitive, sometimes making a centralized cloud a poor match for them. Not to mention the dominant position of the few players that are able to take control of every human activity.

The level of trust and the speed needed by decentralized edge services may be achieved through permissioned blockchains. Problems such as authorization and auditing, two important prerequisites of trust, are naturally solved in permissioned distributed ledgers. 

Furthermore, we do not expect a single global massive blockchain network as a substrate for a decentralized Internet. Open permissionless blockchains may become analogous to the Deep Web/darknet in the current Internet. Anonymous networks designed to hide transactions and information from the rest of the world. While federated permissioned networks will heavily rely on trust and identity, permissionless ones will clearly pivot towards complete anonymity.

Instead of that, we foresee a myriad of permissioned blockchain networks emerging in vertical domains (health, education, energy, automotive, smart cities) with participants across value chains, and possibly located in different legal boundaries (countries, regions, consortiums). In fact, this is already happening in many sectors, because there are  economic incentives in place to build these networks, and these incentives are derived from the business models of the participants, rather than being self-contained in the blockchain (as is the case with cryptocurrencies).

The interoperability of these blockchain islands along with the widespread adoption of decentralized identity services will create major economies of scale in the pioneer countries adopting these technologies. We can present here the analogy of blockchain islands with autonomous systems or even intranets in the existing  Internet. The inherent multiparty structure of these networks, and the variety of them, are perfectly aligned with the original principles of Internet decentralization.

In principle, many of the blockchain networks may be confined to national boundaries, but soon, international blockchain networks will also emerge in different locations. The European Union is a clear candidate for the creation of such multinational networks because of its structure, digital laws (GPDR) and common multiparty connections in every domain. Such networks could become superhubs for a myriad of lesser national networks.

\subsection{BlockChain islands}


We will witness in the coming years how government-backed \textbf{blockchain islands} will emerge around the world in different sectors~\cite{datta-blockchain-in-gov}. In this line, on April 2018, 22 european countries already signed the European Blockchain Partnership (EBP) to cooperate in the European Blockchain Services Infrastructure (EBSI) that will support the delivery of cross-border digital public services, ensuring security and privacy.  Others are following , like the Australian National Blockchain or the Smart Dubai Blockchain.  Let us review some use cases:

\begin{itemize}
\item In \emph{supply chain \& logistics}, distributed ledgers can be used to verify the trade status of products by thoroughly tracking them from their origin to the destination without ever having to explicitly trust any one node in the network
\item In \emph{healthcare}, institutions suffer from an inability to share data securely across platforms. Permissioned blockchains could facilitate hospitals, pharmacies, patients,  clinical research organizations, payers and other parties in the healthcare value chain to share access to their networks without compromising on the data security, privacy and integrity.
\item In \emph{education}, there are strict regulations that the present day educational institutions follow in order to be universally accepted and verified. Validation of academic credentials still remains largely a manual endeavor. Blockchains could help streamline the verification process in academia, thereby reducing fraudulent claims of un-earned educational credits.
\item The \emph{utilities} landscape is evolving into a decentralized and smart power grid, with distributed power generation from both residential and business clients. Simultaneously, utilities are investing in smart meters and upgrading their infrastructures to turn the distributed and smart power grid into a reality. 
As the grid expands and becomes more distributed, establishing trust between stakeholders is fundamental to maintaining effective operations. With blockchains, utilities could provide a trustworthy and secure platform
for distributed grid and smart device usage.
\end{itemize}

What do these business segments have in common? They are all composed by a collection of players or stakeholders that do not fully trust each other. 

In most of the examples, businesses or legal entities are involved, willing to cooperate for the good of their business. They are all capable to dedicate a given amount of reliable resources to such services. This is a completely different scenario with respect to where just end-users are involved, with their limited bandwidth, storage, computation.

While a centralized solution for each individual use case may be feasible, it would be in contradiction with the individual participants' need to control their own data in a local manner, e.g., using tokenization for the blockchain ledger, while controlling the actual content/services. Such an approach would also lack the flexibility of organic evolution and portability of members and services.




\subsection{Challenges}

We foresee interesting challenges in the next years to combine cloud
computing, permissioned blockchains and edge-centric computing in order to
construct the substrate for a decentralized Internet.

We outline the following relevant challenges:
\begin{enumerate}
\item Simplicity and validation of software artifacts: There is an urgent need to reduce
complexity in the blockchain software stack to ensure its massive adoption. The
complexity and frailty of Solidity and Ethereum and other blockchain software
frameworks hinder their adoption. Blockchain software may converge with  Serverless FaaS (Function as a Service) technologies that are
gaining traction in the Cloud thanks to their simplicity and clear APIs. But
more research is needed on software artifacts, programming abstractions, and
software validation and testing on distributed Blockchain code and smart
contracts.
\item Interoperability and open standards: Mass adoption of
Blockchain technologies depends on open standards to communicate across heterogeneous platforms. For example, Hyperledger Composer is offering data integration and connection from Blockchains to existing systems. Hyperledger Indy is also
offering open decentralized identity services that may connect heterogeneous
services.
\item Security and privacy: A never-ending challenge is to provide
security and privacy in a system that is constantly under attack. Encryption as a
service using Blockchains is a major challenge to simplify secure interactions. Decentralized identities and certificates should
seamlessly move between edge devices and distributed blockchains networks. More research is required on encrypted data structures and APIs, sandboxed environments, verification, secure tokens, and zero-knowledge proofs over shared content.
\end{enumerate}

\section{Conclusions}

We reviewed the existing literature to explain the problems of open peer-to-peer networks such as instability, security and system complexity. These problems caused a clear decline of pure peer-to-peer applications in the last years. We also explained how many of these problems still apply to the blockchain open networks such as Bitcoin and Ethereum.

Our claim is that open permissionless blockchain networks cannot become the substrate for a future decentralized Internet. Yet, we believe that more decentralization in the  Internet is still desirable, and can be achieved, answering to the legitimate call for autonomy from  Internet giants and enhanced protection of users' privacy. 


Permissioned blockchains may be the key to this revolution, as they have the potential to deliver dependability and trust in a decentralized manner, instead of relying on any single trusted third party.  Our edge-centric approach leverages the decentralized trust offered by blockchains, co-opting cloud services as a reliable utility, while  preserving users' privacy and control for the Internet of the future.
\newpage






\bibliographystyle{IEEEtran}
\bibliography{IEEEabrv,blockchains_arxiv}

\end{document}